\documentclass{ieeeaccess}

\usepackage{cite}
\usepackage{amsmath,amssymb,amsfonts}
\usepackage{algorithmic}
\usepackage{graphicx}
\usepackage{textcomp}
\usepackage{subcaption}

\usepackage{algorithmic}
\usepackage{braket}
\usepackage{blindtext}
\usepackage{tabularx}

\def\BibTeX{{\rm B\kern-.05em{\sc i\kern-.025em b}\kern-.08em
    T\kern-.1667em\lower.7ex\hbox{E}\kern-.125emX}}

\begin{document}
\history{Date of publication xxxx 00, 0000, date of current version 2022 08.}
\doi{10.1109/TQE.2020.DOI}

\title{Mixed Quantum-Classical Method \\
For Fraud Detection with Quantum \\
Feature Selection }

\author{\uppercase{Michele~Grossi}\authorrefmark{1}, %\IEEEmembership{Fellow, IEEE},
\uppercase{Noelle~Ibrahim\authorrefmark{2},
\uppercase{Voica~Radescu}\authorrefmark{3},
\uppercase{Robert~Loredo}\authorrefmark{4},
\uppercase{Kirsten~Voigt}\authorrefmark{5},
\uppercase{Constantin~von~Altrock}\authorrefmark{6},
and Andreas~Rudnik}\authorrefmark{7}.
%\IEEEmembership{Member, IEEE}
}

\address[1]{European Organization for Nuclear Research (CERN), Geneva 1211, Switzerland (email: michele.grossi@cern.ch)}
\address[2]{IBM Quantum, IBM 3600 Steeles Ave East
Markham, ON L3R 9Z7, CA (email: noel.ibrahim@ibm.com)}
\address[3]{IBM Quantum, IBM Deutschland Research \& Development GmbH, Schönaicher Str. 220, 71032 Böblingen, Germany (email: voica.radescu@ibm.com)}
\address[4]{IBM Quantum, IBM Corp,  1 Alhambra Plaza Suite \#1415
Coral Gables, FL 33134 (email: loredo@us.ibm.com)}
\address[5]{IRIS Analytics GmbH, Klostergut Besselich, 56182 Urbar, Germany (email: kirsten.voigt@iris.de)}
\address[6]{IRIS Analytics GmbH, Klostergut Besselich, 56182 Urbar, Germany (email: constantin.von.altrock@iris.de)}
\address[7]{IRIS Analytics GmbH, Klostergut Besselich, 56182 Urbar, Germany (email: andreas.rudnik@iris.de)}
\tfootnote{M. Grossi, N. Ibrahim,  V. Radescu are primary contributors. List of authors N Ibrahim, V. Radescu, M. Grossi, K. Voigt, and C. Von-Altrock declare that they are authors of patent pending entitled: “Mixed quantum-classical method for fraud detection with Quantum Feature Selection” Nr. P202105918US01 filed on 12/10/2021. We declare that there are no competing interests.}

\markboth
{Author \headeretal: Preparation of Papers for IEEE Transactions on Quantum Engineering}
{Author \headeretal: Preparation of Papers for IEEE Transactions on Quantum Engineering}

%\corresp{Corresponding author: First A. Author (email: author@ boulder.nist.gov).}

\begin{abstract}
This paper presents a first end-to-end application of a Quantum Support Vector Machine (QSVM) algorithm for a classification problem in the financial payment industry using the IBM Safer Payments and IBM Quantum Computers via the Qiskit software stack. Based on real card payment data, a thorough comparison is performed to assess the complementary impact brought in by the current state-of-the-art Quantum Machine Learning algorithms with respect to the Classical Approach. A new method to search for best features is explored using the Quantum Support Vector Machine's feature map characteristics. The results are compared using fraud specific key performance indicators: Accuracy, Recall, and False Positive Rate, extracted from analyses based on human expertise (rule decisions), classical machine learning algorithms (Random Forest, XGBoost) and quantum based machine learning algorithms using QSVM. In addition, a hybrid classical-quantum approach is explored by using an ensemble model that combines classical and quantum algorithms to better improve the fraud prevention decision.  We found, as expected, that the results highly depend on feature selections and algorithms that are used to select them. The QSVM provides a complementary exploration of the feature space which led to an improved accuracy of the mixed quantum-classical method for fraud detection, on a drastically reduced data set to fit current state of Quantum Hardware.
\end{abstract}


\titlepgskip=-15pt

\maketitle
\section{Introduction}
Over the past few years, the financial industry has seen a substantial growth in innovation, particularly in the field of AI/ML with respect to the payment industry in an effort to keep fraud losses contained \cite{ryll2020transforming}. 
The current challenges are those of finding the balance between the false positives where, if too common, could serve as a negative impact to a client's experience \cite{VOROBYEV2022102786} and minimizing the monetary loss incurring by fraudulent transactions. 
%In particular with the statistically skewed distribution of payment fraud that exists today. 
%Assuming that 1 out of 1000 transactions is fraudulent, a false positive rate of only 1\% implies 10 transactions are wrongfully declined \red{NEED ref for this claim.} \cite{?}.
Yet criminals are also constantly increasing their capabilities to deploy ever more complex fraud schemes at a rate difficult to keep up.
Many have started using AI/ML to augment the efficacy of their attacks \cite{yeoh2019artificial}. 
The payment industry defends itself in multiple ways: more data from more sources is used, more behavioral features are extracted as inputs to the AI/ML models and better machine learning models. This is an area where quantum computing could provide a disruptive improvement, in particular by identifying features that lead to more accurate classification.
%In related application areas, where predictive models must represent multidimensional, highly non-linear, and combinatorial problems, quantum computing has provided fast and efficient solutions. 

Quantum machine learning is an active field of research which seeks to take advantage of the capabilities of both quantum computers and machine learning techniques, adapting the latter to the strengths of the current state-of-the-art in quantum computing. There are many examples that illustrate how quantum computing can be used to train models \cite{abdelgaber2020overview,schuld2018supervised} and possibly enhance machine learning models such as quantum support vector machines (QSVM) \cite{havlivcek2019supervised,rebentrost2014quantum}, quantum classifiers (QC) \cite{yano2020efficient}, and quantum neural networks (QNN) \cite{abbas2021power}.  
%\ref{prior:ref1, prior:ref2, prior:ref3, prior:ref4}.
%\red{TODO: (DONE - Robert to include some references to current state-of-the-art in QML that align with fraud detection or other financial services).}
Much work has been conducted on synthetic and publicly available datasets from various domains such as drug discovery \cite{batra2021quantum}, image classification \cite{Kerenidis_2020}, and computational sciences \cite{ciliberto2018quantum}. Comparisons have been made to the classical counterparts of the available quantum machine learning algorithms \cite{stein2021hybrid}. In addition, when synthetic data is used for machine learning experiments, there have been provable advantages shown involving synthetic data sets when there is a lack of necessary data \cite{lopez2012money,abufadda2021survey}.

In this work we investigate the impact of quantum feature selection techniques versus classical feature selection techniques on the performance of the quantum machine learning classifier. 
We consider that prefacing the classical feature selection to the application of a quantum machine learning may eliminate some or all of the complex nuances in the relationships between features and outcomes that quantum machine learning methods are thought to be able to detect. Finally, we compare the performance of Quantum Support Vector Machines to state of the art methods in fraud detection such as Random Forest and XGBoost, using a "real-world" data set of card payment transactions with real fraud marks. We also introduce the concept of mixed quantum/classical machine learning ensembles, and test these against the model performance of the purely classical and purely quantum approaches. 

\subsection{Methodology}
The three industry methods are being analyzed in this paper using same initial data set:
\begin{enumerate}
    \item Domain expert created decision rules-based model (no machine learning)
    \item State-of-the-art type AI/ML using boosted trees (Random Forest, XGBoost)
    \item Quantum Support Vector Machine (QSVM) type model
\end{enumerate}
%To compare modelling capabilities of the three modelling methods, we have used the same standard set of behavioral features that were devised by the domain expert from experience. 
As experimentation and potentially later real-world deployment platform, we are using IBM Safer Payments software product. IBM Safer Payments is unique in providing real-time and offline monitoring of payment transactions with internally and externally AI machine-learning models. We have first loaded the transaction data and computed the behavioral features. We then created the domain expert-based additional features within IBM Safer Payments. We exported the training data for 2. and 3. directly from IBM Safer Payments to assure compatibility of the model with input data. This is an important aspect when discussing integration. 
%and providing the tools to construct or improve those models whenever needed to stop new fraud.  
%The results of the externally trained models were put back into Safer Payments for analysis and comparison. 
If the QSVM model is to be used with payment processor’s production system, the integration with the IBM Safer Payments product is feasible due to the external model import capabilities already built in the product. However, additional considerations related to latency requirements should be accounted when discussing integration. This is not within the scope of this paper.
%For the real-life usage of our results, latencies 
%we can also import the models as runtime code. Thus, real-life usage of our results is immediately provided. 

\subsection{Payment Fraud Prevention KPIs}
Payment fraud prevention relies on two specific KPIs: (monetary) \textit{it rate}  and \textit{false alarm ratio}.
The hit rate, typically reported as a percentage, is the total amount of all transactions that were declined by the fraud prevention system that later were confirmed to have been fraudulent, divided by the total amount of transactions that were later confirmed to have been fraudulent. 
The false alarm ratio is typically given as the ratio of false alerts to true alerts. Thus, if the model created $10$ false alerts for every true alert, it has a false alarm ratio of $10$:$1$. Each false alert causes disruption of a customer’s payment, and triggers potential manual interaction with the customer, both which are mostly independent of the amount of the transaction.
%In other words, it is the “loss savings rate” of the model. The hit rate is typically given as the ratio of false alerts to true alerts.

%Thus, if the model created 10 false alerts for every true alert, it has a false/positive rate of 10:1. Notice that this is different from a “standard ROC*” view. Mostly because for the loss savings (hit rate), the monetary amount is of prime importance rather than the number of transactions. To the fraud loss savings, it is indifferent if there is one fraudulent transaction of \$100 prevented or two fraudulent transactions of \$50 each. The false/alerts, however, are considered on a per-event basis. Each false alert causes disruption of a customer’s payment, and triggers potential manual interaction with the customer, both which are mostly independent of the value of the transaction. Thus,in the remainder of this paper, we will both use these“industry KPIs” and standard ROC* representation of the results. 
When invoking Machine Learning Classifiers for payment fraud prevention, these are for a binary classification (fraud vs non-fraud). A statistical measure of a model is given by \textit{Accuracy}, the number of classifications a model correctly predicts divided by the total number of predictions made. The accuracy KPI is meaningful only for a balanced class data set. 
A better diagnostic of a binary classifier performance is through a Receiver Operating Characteristic curve, or ROC curve. \textit{Area Under the ROC curve (AUC)}  is one of the most important evaluation metrics for checking any classification model’s performance. AUC represents the degree or measure of separability. We have adapted it to align better with the financial KPIs, so that the \em{x} \rm axis is the false alarm ratio, therefore throughout the text we refer to this curve as modified ROC curve or ROC*.

Machine Learning Classifiers are used for generating a score. This score could be on an ordinal scale, or it could be representing the predicted probability of the current transaction turning out to be fraudulent later. Since the real-time decision can only be to either decline or not to decline a transaction, usually a threshold is applied to the score to make this decision. 

\section{Input Data Set}
We are using a data set of real-world payment transactions that comes from a European cross-border processing portfolio and consists of about $80$\% debit and $20$\% credit card transactions. This data set contains a total of $2.4$ million payment transactions. Each transaction is flagged as fraud or non-fraud, with a total of \~3k transactions marked as fraudulent in the data set. Importantly, the transaction data can be enriched with customer reference data and with features built on the fly, as described in the following subsections. 

\subsection{Transaction and Customer Data}
The data set we work with has only $12$ input attributes from transaction data, with additional $2$ attributes from demographic data, the remaining ones are engineered through discovery techniques, as described next. 
%A typical transaction input attributes are listed in the in the Appendix, in table \ref{tab:features_1}. 
It is usually possible to enrich the transaction information with demographic data available within the financial institution for the card holder that initiated the payment, usually referred to as "customer reference data" or "masterdata". Examples of reference data: customer data linked to an account or card number, additional information related to merchants, supporting technical data such as the countries that correspond to card number ranges (BIN/IIN), IP addresses, etc.

\subsection{Engineered Features}
An important ingredient to a model is feature discovery. Engineered features such as behavioral profiles formed from the transaction inputs encapsulate meaningful information for classification problems. Profiles provide aggregated counts of totals and transaction frequencies over calendar periods or pre-defined time windows for every customer or card number that is indexed in the database. Since they encapsulate a history of a transaction fulfilling the counting conditions and certain patterns, these features provide a strong discriminating power between a fraudulent and non-fraudulent transaction. 

In total, before invoking any pre-processing of data, we have $48$ attributes. 
These attributes are of various data types: categorical, string, integers, etc. Handling categorical data type is posing some challenges for machine learning classifiers, and it will require treatment such as one hot encoding techniques and/or clustering of the relevant values. 

\subsection{Datasets for Use Cases}
The aim is to compare the impact of different methodologies by analyzing the same input data. However, different methodologies may have different limitations. For example, human expertise and rule generator do not necessary require a balanced data set in terms of fraudulent vs genuine records, whereas machine learning methods (classical or quantum) require balancing the set via under-sampling methods. Moreover, when using Quantum Machine Learning (QML), Quantum Hardware is limited in number of qubits and error rates, one needs to reduce data dimensionality considerably while maximally preserving the accuracy of the model.

Therefore, we conducted this study using three distinct data references that require different levels of pre-processing for each of the analyzed use cases:
\begin{enumerate}
    \item Full dimensionality of a data set:  only cleaned from redundant data and split into train and test. This data set can be used by Rule Generator assisted by a fraud subject matter expert, even if it's highly imbalanced by the number of genuine records relative to the fraudulent ones.
    \item Balanced data set, with genuine transactions randomly under-sampled (there are various methods to achieve this), together with treatment to handle the categorical data type. This data set then can be optimally used by Classical Machine Learning.
    \item Drastically reduced data samples which will require multiple trials to avoid bias due to heavy under-sampling. In addition, further normalization of all input data is applied to ease the translation to Quantum Feature Mapping. This dataset is then used as for a direct comparison of the classical and quantum methods.
\end{enumerate}

%\section{Feature Selection}
%MISSING SECTION

\section{Classical Fraud Detection Model}
%Creating the best possible decision model is essential to achieving maximum fraud prevention performance. IBM Safer Payments contains fully integrated modeling tools that are easy to use by fraud experts. The decision model is used by the real-time service to determine whether a transaction could be fraudulent or not.
The process of creating decision models in fraud prevention systems such as IBM Safer Payments is invoked either via a conventional rule-based approach or using machine learning techniques.%involves identification of input attributes used for indexing the dimension to create aggregated features. Once the data is enriched with these features, a decision model can be 

\subsection{Human and Interactive expert method}

For the assisted and automatic model generation, a hybrid-logic rule generating algorithm is used.
The algorithm creates rule sets condition by condition and rule by rule. The assisted model generation proposes the next generation step where the expert can then either accept the suggestion,  modify the parameter of the proposed condition, or not follow the proposal at all and select an own condition.
The automatic model generation assumes an acceptance of each proposal and stops only if a defined stop criteria is reached.
A model generation uses statistical analysis to discover fraud patterns that can be aided by a human fraud expert.

The Rule Generator is based on a deterministic algorithm to search the data set for specific fraud patterns.  The Rule Generator can account for categorical attributes by default, and all attributes can be included without treatment for categorical attributes. No under-sampling is necessarily needed, because the human method and rule generator are defining behavioral patterns with the focus to catch a fraudulent transaction. The Rule Generator model is trained on a training set and validated on a test set. 
%The full data set of $2.4$mil records is split into training of $1.5$mil records and verification $0.9$mil records. 

There are various ways to split data. However, in order to mimic the real-life situation where training is done on past data and prediction applied on new incoming transaction, data has been split chronologically as follows: the first $1.5$million records are used for training and the remaining records for testing. The count of records in Fraud and Genuine is shown in Table \ref{tab:original} (1:1000 Imbalanced data).
%\begin{table}[]
%    \centering
%    \begin{tabular}{c|c|c}
%        Train & Label & Count  \\
%        \hline
%        &0 & 1498434\\
%        &1 & 1566\\
%        \hline
%        Test & Label & Count  \\
%        \hline
%        &0 & 898255\\
%        &1 & 1650\\
%        \hline
%    \end{tabular}
%    \caption{Splitting of data set into training and test. This leads to 1:1000 Imbalanced %set}
%    \label{tab:original}
%\end{table}

\begin{figure}[t]
  \centering
  \includegraphics[width=0.45\textwidth]{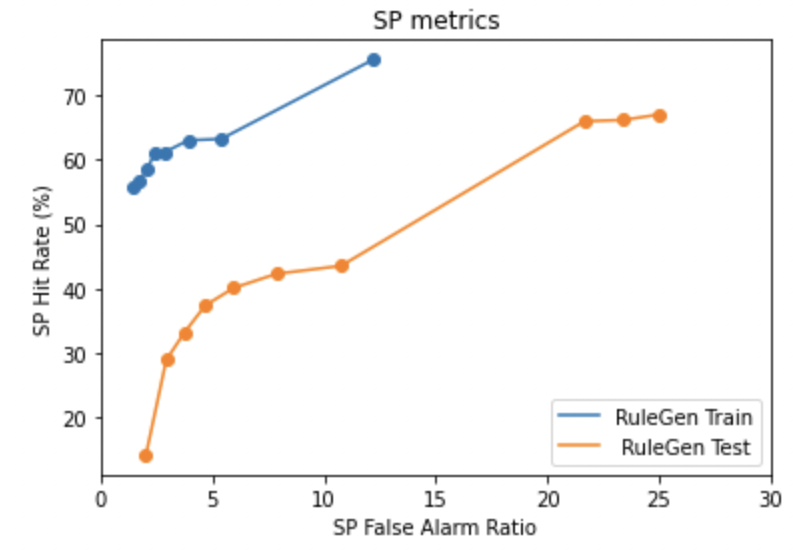}
  \caption{Results based on Rule generator using a complete data set that is split into training (blue) and verification (orange). The metrics are Hit Rate and False Alarm Ratio. This representation of the KPIs is often referred to as a modified ROC* curve, with the False Alarm Ratio on the $x$ axis (ratio False Positives \/ True Positives. }
  \label{fig:RuleGen}
\end{figure}

The results obtained in terms of fraud prevention KPI are Hit Rate and False Alarm Ratio and shown in Figure \ref{fig:RuleGen}, often referred to as a modified Receiver Operating Characteristic (ROC) curve. The modified ROC* curve is the primary metric to understand the performance of a model. 
 
The KPIs of the model can be as good as $65$\% hit rate but with the penalty of $25$ non-fraudulent payments intercepted for each one fraudulent payment intercepted, or with 30\% hit rate with only $5$ non-fraudulent payments intercepted for each one fraudulent payment intercepted. This is a trade-off decision to be taken by either a processor or a bank. The train curve sits considerably above the test curve which is an indication that Rule Generator method is prone to over fit. 
 
 \subsection{Classical Machine Learning Classifiers}
Payment Fraud Detection is commonly using supervised machine learning classifiers, where historical data has fraud marks either detected by a case investigator or reported by an affected customer and has ideally been caught by the fraud prevention models before it happened. 
Examples of Supervised Learning: Regression, Decision Tree, Random Forest, SVM,  Logistic Regression etc.
For this analysis, we've explored decision trees-based models such as XGBoost and Random Forest, as they are known to outperform other supervised learning classifiers.

\subsubsection{Data Preparation for Classical Classifiers}
There are mainly two drivers for data preparation for a classical classifier: balance the data set and convert data types to numeric values. 
%The rationale for the former case is that accuracy is a measure of how rightfully a classifier classified a record. W
The following pre-processing steps have been applied to data before passing it to a classical classifier, apart from the under-sampling: 
\label{DataPrep}
\begin{enumerate}
    \item Removal of highly correlated features (duplication of information)
    \item Treatment for categorical data types: classify top categories where most fraud occurred in historical data and use them as separate features. This step has increased the number of features from $48$ to $69$. 
    %The new list of attributes in the Appendix.
    \item Split of the data into “training” and “test” sets for use when training the models.
    \item Treatment for imbalanced data, where there is much more genuine than fraudulent transactions: achieved by under-sample genuine by larger fraction than fraud with the aim to preserve all fraud marks. Finding the right balance is an art, we ran $5$ random trials.
\end{enumerate}
%\begin{table}[]
%    \centering
%    \begin{tabular}{c|c|c}
%        Train & Label & Count  \\
%        \hline
%        f=0 &0 & 1498434\\
%        f=0 &1 & 1566\\
%        \hline
%        Test & Label & Count  \\
%        \hline
%        f=0 &0 & 898255\\
%        f=0 &1 & 1650\\
%        \hline
%    \end{tabular}
%    \caption{Splitting of data set into training and test. This leads to 1:100 Imbalanced set}
%    \label{tab:original}
%\end{table}

%The attributes that were selected for finding the optimal hit rate and false positives can be found in the Appendix, \ref{tab:xgbaccuracy}.

We first started with a complete data set which has been split into test and training, as in Table \ref{tab:original}  and then, on reduced data set that has been used for the Quantum Machine Learning part as well. 
\begin{table}[]
    \centering
    \begin{tabular}{c|c|c|c|c}
        Set & Label & Count & Train & Test \\
        \hline
        \hline
%        Original &0 & 2396689 & 191733 & 479354\ \\
%                 &1 & 3216 & 2589 & 627 \\
        Original &0 & 2396689 & 1498434 & 898255 \\
                 &1 & 3216 & 1566 & 1650 \\          
        \hline
        Balanced  &0 & 1505  & 984 & 521\\
                  &1 & 993 &  515 &  478\\
       \hline
        Reduced  &0 &  366& 262 & 104\\
                 &1 &  232 & 137& 95\\
        \hline
        \hline
    \end{tabular}
    \caption{Original Data Set Count, Balanced Data Set Count and Drastically Reduced Data set count, overall and split into test and train. }
    \label{tab:original}
\end{table}

\subsubsection{Classical Machine Learning with Original Data Sets}

%Like before for the Rule Generator, the supervised machine learning models require a split of data into a training and a testing data set. 

%For simplicity, we have dropped the categorical attributes when we used XGBoost and Random Forest. The results based on the full data set can be found in Table \ref{tab:xgbaccuracy}
%\begin{table}[]
%    \centering
%    \small
%    \begin{tabular}{p{2.4cm}|p{2.cm}|p{2.3cm}}
%XGboost &Accuracy (Train)& 0.999\\
%XGBoost &AUC Score (Train)& 0.894424\\
%\hline
%XGBoost &Accuracy (Test)& 0.9982\\
%XGBBoost &AUC Score (Test)&  0.845026\\
%\hline
%\hline
%RF &Accuracy (Train)& 0.999 \\
%RF& AUC Score (Train)& 0.848127 \\
%\hline
%RF & Accuracy (Test)& 0.9982 \\
%RF & AUC Score (Test)& 0.834642 \\
%    \end{tabular}
%    \caption{Accuracy and AUC results for XGBoost and RF}
%    \label{tab:xgbaccuracy}
%\end{table}

We used \textit{XGBoost CV}  package for tuning the model's hyper parameters to find the best number of estimators, max\_depth, min\_child, in a non-exhaustive iterative approach to find optimal values of these input parameters.
%The best resulting model was found for the parameters provided in Table \ref{tab:xgbparam}.
%\begin{table}[]
%    \centering
%    \small
%    \begin{tabular}{l|l}
%    Parameter Value\\
%    \hline 
%colsample\_bytree=0.8,\\ 
%learning\_rate=0.01,\\
%max\_depth=5,\\
%min\_child\_weight=2, \\
%n\_estimators=136,\\
%nthread=3,\\
%seed=27,\\
%subsample=0.8,\\
% \end{tabular}
%\caption{XGB Parameters}
%%   \hline 
%\label{tab:xgbparam}
%\end{table}

Similarly, for tuning Random Forest parameters we used \textit{RandomizedSearchCV} package, where we identified the optimal number of trees needed in random forest, number of features to consider at every split, maximum number of levels in a tree, minimum number of samples required to split a node, and minimum number of samples required at each leaf node.
We used Random search of parameters, using $3$-fold cross-validation, and searched across $10$ different combinations. 
The results of these two fits in terms of Accuracy and AUC performance parameters are shown in Table \ref{tab:modelaccuracy}.
\begin{table}[ht]
    \centering
    \small
    \begin{tabular}{p{2.4cm}|p{2.cm}|p{2.3cm}}
    KPI & XGBoost & Random Forest\\
    \hline
    \hline
Accuracy (Train)& 0.998 & 0.999\\
AUC (Train)& 0.813 & 0.999\\
\hline
Accuracy (Test)& 0.998 & 0.998\\
AUC (Test)&  0.824 & 0.818\\
\hline
\hline 
    \end{tabular}
    \caption{Accuracy and AUC results for XGBoost and Random Forest using Original Data, without under sampling}
    \label{tab:modelaccuracy}
\end{table}

Interesting to observe is the list of feature importance ordered by the classical classifiers in Table \ref{tab:imp}. Each classifier has a different preference for the order of importance (impact) of the features. This is why choosing a different methodology for machine learning provides a different view of the feature space. This observation also prompted us to study the feature selection using QML where Quantum Machine Learning can complement classical methodology to improve the Fraud KPIs.

\begin{table}[]
    \centering
    \small
    \begin{tabular}{l|l}
     XGBoost & Random Forest\\
     \hline
     \hline
%    C\_4h\_Shopping, &  Cal\_MCC\_FREQ,\\
%    Cal\_MCC\_FREQ, & Amount, \\
%    Cal\_Last4d\_TotAmt, &  Cal\_Last4d\_TotAmt,\\
%    E\_TimeLastATMwoInclCurr, & Cal\_Yesterday\_TotAmt, \\
%    Amount, &  Cal\_Card\_FREQ\\
%    C\_3h\_Declined, &  C\_4h\_Shopping\\
%    C\_24h\_FREQ &  E\_TimeLastATMwoInclCurr\\
    F\_10 & F\_16 \\
    F\_16 & F\_0 \\
    F\_15 & F\_15 \\
    F\_0 & F\_14 \\
    F\_7 & F\_10 \\
    F\_3 & F\_19 \\
    \hline
    \hline
 \end{tabular}
    \caption{Ordered feature Importance for XGBoost and Random Forest.}
    \label{tab:imp}
\end{table}

\begin{figure}[t]
  \centering
  \includegraphics[width=0.45\textwidth]{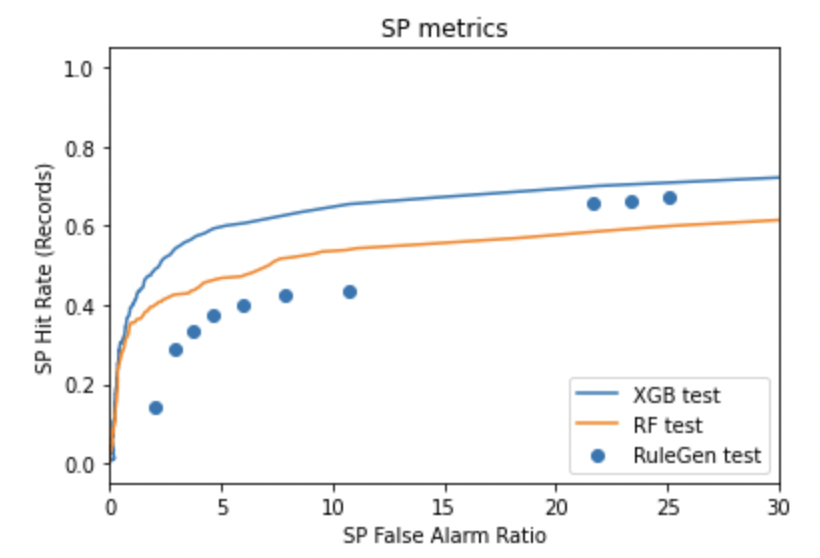}
  \caption{Results based on XGBoost, Random Forest and Rule Generator using a complete data set without categorical attributes that is split into training and verification. The metrics are Hit Rate and False Alarm Ratio.}
  \label{fig:classical}
\end{figure}

To compare them with the Rule Generator, which uses a different KPI metric, we have also looked at the Hit Rate vs False Alarms, as defined in the Safer Payments product.  The figure \ref{fig:classical} overlays the modified ROC* curves from different machine learning models versus Rule Generator.

Figure \ref{fig:classical} shows the relation between Hit Rate (1 means 100\% and 0 is 0\% correctly identified fraudulent payments) on the y axis, while x axis represents the false alarm rate, meaning disturbed clients for each true fraud. 
The ideal case is a Hit Rate of $100$\% and a False Alarm Rate of $0$. As you can see in this Figure, reaching a Hit Rate of $50$\% of intercepted fraud would cause the False Alarm rate to rise to  more than $1:50$. This result, in fact, is better than using Rule Generator, capturing more fraudulent records than a model trained via Rule Generator. Similarly, XGBoost seems to outperform the Random Forest model for this data set. However, the Hit Rate from XGBoost and Random Forest is on a record count basis, while from the Rule Generator is per amount value basis.

%\begin{table}[]
%    \centering
%    \small
%    \begin{tabular}{p{5.4cm}|p{8.3cm}}
%Feature& Score\\
%\hline
%Cal\_MCC\_FREQ &   0.130700   \\
%Amount & 0.111093 \\                C\_24h\_SUM &  0.089760   \\   
%Cal\_Last4d\_TotAmt & 0.082906  \\      Cal\_Yesterday\_TotAmt &     %0.074406 \\
%C\_4h\_Shopping & 0.060730   \\       
%Cal\_Card\_FREQ & 0.057459   \\
% \end{tabular}
%    \caption{Feature Importance for Random Forest.}
%    \label{tab:rfimp}
%\end{table}
%\begin{figure*}[!t]
%\centering
%\subfloat[Case I]{\includegraphics[width=2.5in]{box}%
%\label{fig_first_case}}
%\hfil
%\subfloat[Case II]{\includegraphics[width=2.5in]{box}%
%\label{fig_second_case}}
%\caption{Simulation results for the network.}
%\label{fig_sim}
%\end{figure*}

 \subsubsection{Classical Machine Learning with Balanced Data Sets}

\begin{figure}[t]
  \centering
  \includegraphics[width=0.4\textwidth]{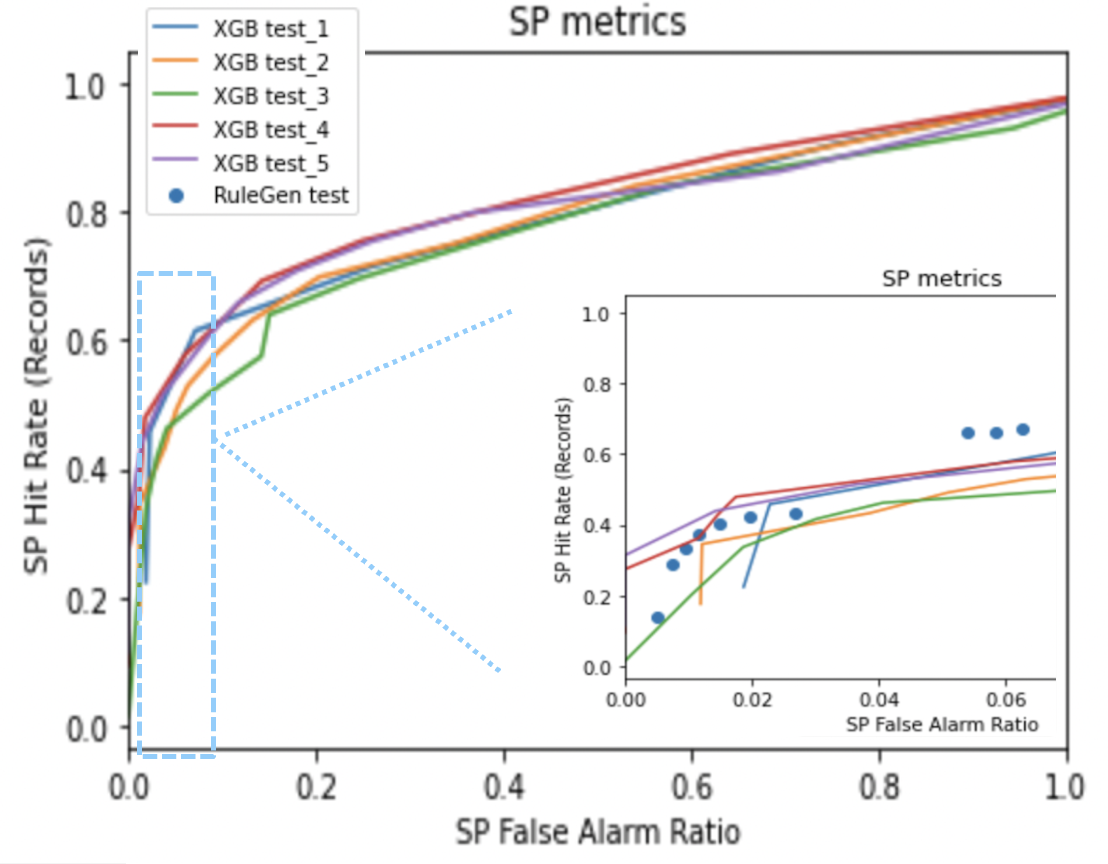}
  \caption{Results based on XGBoost multiple trials. The X axis is false alarms, however the range is equivalent to a factor of ~$500$ more, due to the under-sampling. The snippet is a zoomed region of the plot that would be equivalent to real fraud alarm range of $(0,100)$ and it is directly compared with results from the Rule Generator.}
  \label{fig:XGBMultiple Trials}
\end{figure}

For this part, to balance the set, we massively under-sampled data by $1:1000$ for the genuine transactions, preserving a third of the fraudulent records. The train set has $1500$ records, while the test $1000$, as shown in Table \ref{tab:original}. To minimize biasing effects from the strong under-sampling, we have run the split in $5$ separate trials, preserving the same fraudulent to genuine transaction ratio. The results are captured in Table \ref{tab:accuracy} and an average is computed for reporting KPIs. We observe only small deviations across the trial runs for the KPIs: accuracy, AUC, and the dependency between hit rate vs false alarm rates for the test sample, as seen in the modified ROC* curves Fig \ref{fig:XGBMultiple Trials}. Notice that values on the $x$ axis are affected by the under-sampling and should be scaled by the under-sampling factor of $~500$ to be comparable with the ROC* curves from the original data set. The zoom in the plot is aimed to help guide the eye to run that comparison: the "real" false alarm ratios in the range of $0-100$ correspond to a range of $20-60\%$ hit rate for catching fraudulent records. Therefore, this is comparable with the previous results that did not use sampling, as it is shown in the zoomed plot, where dotted lines correspond to the results from Rule Generator using test data. To be noted that Rule Generator's hit rate has monetary value, while the hit rate from classical machine learning models is based on record count. 
This is an important validation of under sampling procedure, having in view that we can only use a significantly reduced data sets with the current state of quantum hardware. 

\begin{table}[]
    \centering
    \tiny
    \begin{tabular}{l|l|l|l|l|l|l}
    KPI & XGB1 & XGB2 & XGB3 & XGB4 & XGB5  & Average\\
    \hline
    \hline 
%Accuracy (400 Train)& 0.809 & 0.756 & 0.787  & 0.769 & 0.812 & 0.787 \pm 0.021 \\
%AUC(400 Train)& 0.870 & 0.758 & 0.812 & 0.826 & 0.838 & 0.821 \pm 0.036\\
Accuracy (Test)& $0.785$ & $0.774$ & $0.767$  & $0.783$ & $0.796$ & $0.781 \pm 0.010$ \\
AUC(Test)& $0.832$ & $0.837$ & $0.823$ & $0.852$ & $0.845$ & $0.834 \pm 0.010$\\
\hline
%Accuracy (200 Test)& 0.784  & 0.778 & 0.738 & 0.785 & 0.794 & 0.769 \pm 0.020 \\
%AUC (200 Test)&  0.835  & 0.824 & 0.807 & 0.842 & 0.851 & 0.832\pm 0.015 \\
\hline
\hline
    \end{tabular}
    \caption{Accuracy and AUC results for XGBoost and Random Forest using Balanced Data Set, with under-sampling, therefore we took 5 trials.}
    \label{tab:accuracy}
\end{table}

\section{Quantum Machine Learning for Fraud Detection}

There are various quantum machine learning approaches for classification problems \cite{havlivcek2019supervised,stein2021hybrid,orus2019quantum,el2018forecasting}. In this paper we are focused primarily on the Quantum Support Vector Machine (QSVM) approach. 
The search for an increasingly high-performance model is the basis of every research project, and exploring the usage of quantum algorithms is a promising approach. The ultimate goal is to find a quantum kernel that provides an advantage in the classification of real world data by improving a metrics like the classification accuracy. A general recipe for building these kernels is not yet available, except in specific cases such as the definition of class of quantum kernels related to covariant quantum measurements such the one introduced in \cite{glick2021covariant}, applicable to group-structured data. Those kernels can be optimized using a technique called \textit{kernel alignment}.

The motivation for this work is to leverage the QSVM approach in two parts in order to optimize the fraud detection system. The first is to determine which of the many features available should be selected to reduce the dimensionality of the data set for running the experiment on a quantum system.  The second is to derive the fraud KPIs from a quantum machine learning model. We used Qiskit \cite{Qiskit}  quantum software package  
for this work.

\subsection{Quantum Feature Importance Selection Algorithm}%Michele
 The main challenge in the near-term quantum devices is the limited number of qubits. According to the data encoding procedure adopted, which in this case is the one introduced in the QSVM paper \cite{havlivcek2019supervised} where each qubit is associated with a feature, we need to reduce the feature dimensionality of the original dataset to be managed on a real quantum device. 
 
Not only the number of qubits, hence the number of features selected is important, but also is the number of records used for the training sample.  Therefore, using under-sampling techniques to scale down data is an important pre-requisite. All data values are also normalized to the interval $[-1,1]$ using MinMaxScaler package as a more convenient choice for quantum processing of those data mapped as angle rotations $[0,2\pi]$ 
%This is even more true when the execution of the algorithm on real hardware represents just the final step of  standard approaches adopted in the vast majority of cases.

For the feature selection, we started by evaluating several classic methods to reduce data dimensionality for the number of features: from the classical Principal Component Analysis (PCA) method types to the  feature importance extraction from XGBoost or Random Forest on full data set of $2.4$ mil records. 

\begin{figure}[t]
  \centering
  \includegraphics[width=0.4\textwidth]{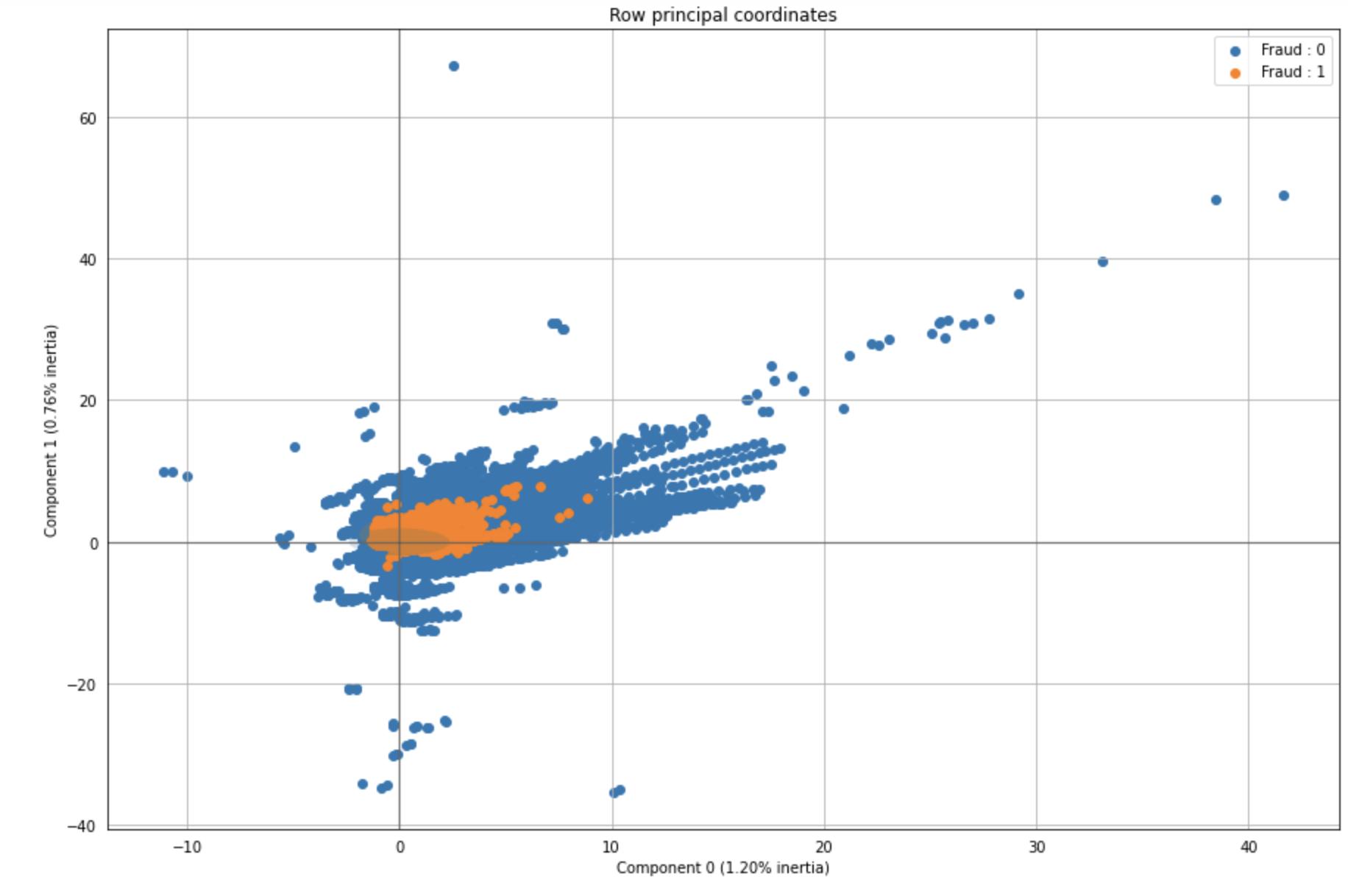}
  \caption{The relation between the components from the FAMD method. A total overlap is observed which limits the FAMD approach in using it for data feature reduction.}
  \label{fig:famd}
\end{figure}

The data used for payment fraud prevention is mostly composed of binary or categorical data types, while the PCA method is designed for continuous variables and hence we could not use it. We experimented with Factorial Analysis of Mixed Data  (FAMD) method,  which works for a mix of categorical and numerical variables. However,  for this data set the method did not show any discrimination power between its reduced variables, displaying a total overlap as shown in Fig. \ref{fig:famd} for the first two components. 

As observed in Table \ref{tab:imp}, different features are preferred by different Machine Learning Classifiers. 
At the time of writing, there is no inbuilt feature importance method for QSVM, while when using features importance extracted by classical classifiers bias the outcome and may undermine the full performance of a QML model, in this case QSVM. 
Therefore, instead of approaching variable reduction through purely classical techniques (which is best adapted to the classical machine learning), we developed a quantum algorithm that would allow use of quantum feature map and quantum kernels to determine best features.   

The Quantum Feature Map $\rho(\cdot):= \ket{\psi(\cdot)}\bra{\psi(\cdot)}$  embeds a data point to a quantum state, so that we can build the classification model, in particular the kernel function that measures the similarity between two data points $x,\,y\in\mathcal{I}$ in the Hilbert space, with respect to the Hilbert-Schmidt inner product as
\begin{align*}
k(x_i,x):= \phi(x_i)^\dagger\cdot\phi(x)=\text{tr}[\rho(x)\rho(y)]\equiv\left|\braket{\psi(x)|\psi(y)}\right|^2 \\
\equiv \left|{\braket{0|U(x)^\dagger U(y)|0}}\right|^2,
\end{align*}
where the quantum feature map is precisely the density matrix $\rho(\cdot)$, $U(\cdot)$ corresponds to a data encoding quantum circuit that represents the quantum feature map and $\ket{0}:=\ket{0}^{\otimes n}$. In our case the ZZ Quantum Feature Map is defined as $ U_{\phi(x)} = \exp(i x_0 Z_0 + i x_1 Z_1 + i (\pi - x_0) (\pi - x_1) Z_0 Z_1)$, where $0,1$ are qubit indexes.
In terms of circuit representation, it is given by Hadamard gates at the beginning and in the middle of the circuit to create quantum interference, followed by a single qubit rotation around the Z axis to encode each feature, and eventually a second order expansion to account for interactions in the data, given by another single qubit rotations of generally the product of two features sandwiched between two controlled $2$ qubit gate. As an illustration,  a quantum feature map using $3$ qubits  is represented in Fig.\ref{fig:depth}.
The minimization of the objective function is realized on a classical device, while the  kernel values are sampled from a quantum computer. 
With our training and testing data sets prepared, we proceeded by setting up the \textit{QuantumKernel} class to calculate a kernel matrix using the \textit{ZZFeatureMap}.

Firstly, the application of QSVM method has been tested and ran on a quantum simulator, under ideal condition, namely with a $state\_vector$ simulator, then in a more realistic scenario with a noisy simulator and eventually on the real device together with error mitigation. This is particular convenient because the first iterations are quite expensive in terms of calculation since the total number of permutations range from roughly half a million to thousands.

\begin{figure}[t]
  \centering
  \includegraphics[width=0.48\textwidth]{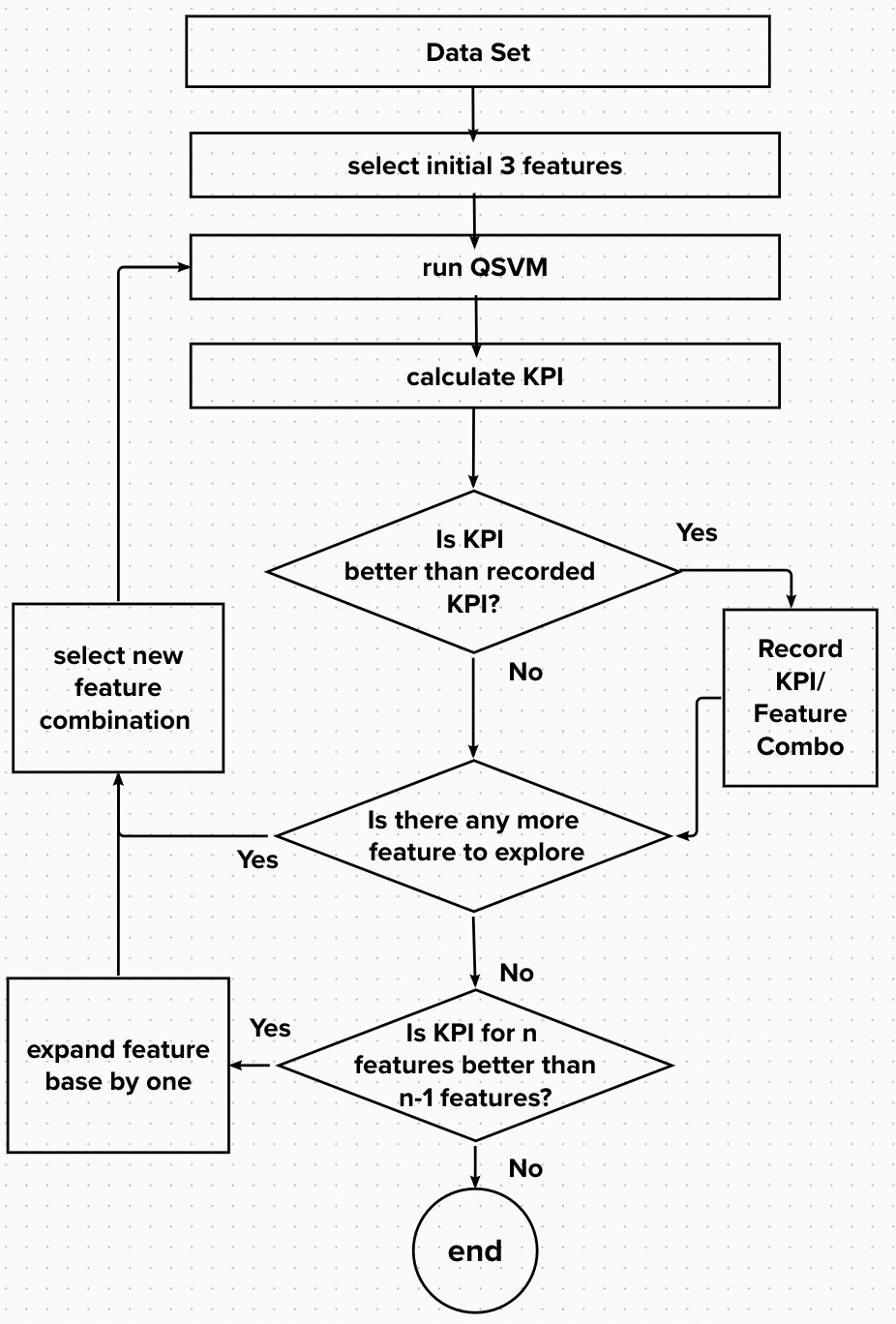}
  \caption{Schematic representation of the quantum algorithm feature selection to exploit the usage of quantum computing in the context of fraud detection.}
  \label{fig:framework}
\end{figure}

We have been inspired by the classic Feed Forward Feature Selection (FFFS) based on AUC or Accuracy as statistical metrics. In this way we can iteratively select an increasing number of features in the problem, i.e. starting from $3$ out of total of $69$.  This quantum approach is integrated and is part of the overall framework defined in Fig.\ref{fig:framework} to approach the problem of fraud detection.
%\subsubsection{Algorithm Description}

Given a dataset $\mathcal{X}$ of dimension $n \times m $, where $n$ is the total number of sample (transaction) and $m$ is the total number of features, the algorithm does a permutation over all the possible  combination of $p$ (starting from 3) features over $m$. For each combination a quantum classifier is defined, trained, tested and the accuracy and the AUC is stored. At the end of the procedure, the best model based on the accuracy as key performance indicator (for future work, one could consider different KPI as the discriminator) and therefore the best $3$ features  out of $69$ are chosen as a baseline for the next model iteration. This leads to exploration of few thousands of combinations (where repetition is not allowed). At this point, the fourth feature is chosen after a permutation over all the remaining features together with the previously selected (only $66$ features are explored). The process can be iterated adding one feature for each permutation cycle up to the desired number of featured, preferably when the improvement saturates. This number can be chosen as a trade-off between the maximum number of available qubit and the total accuracy obtained in the iteration.

\begin{figure}[t]
  \centering
  \includegraphics[width=0.24\textwidth]{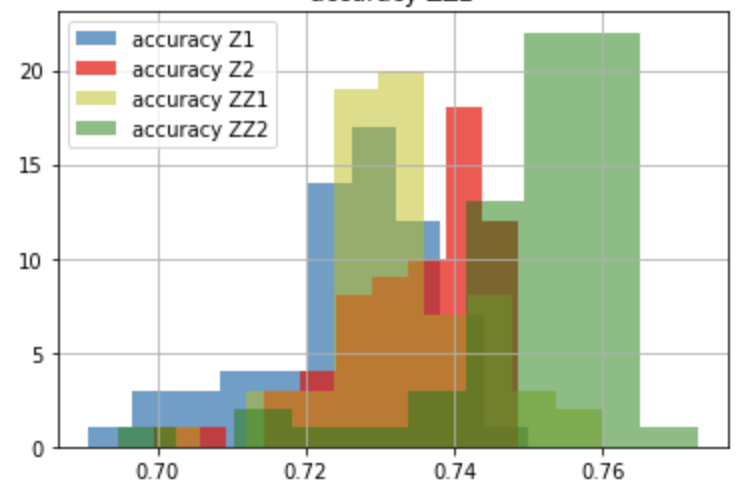}
  \includegraphics[width=0.24\textwidth]{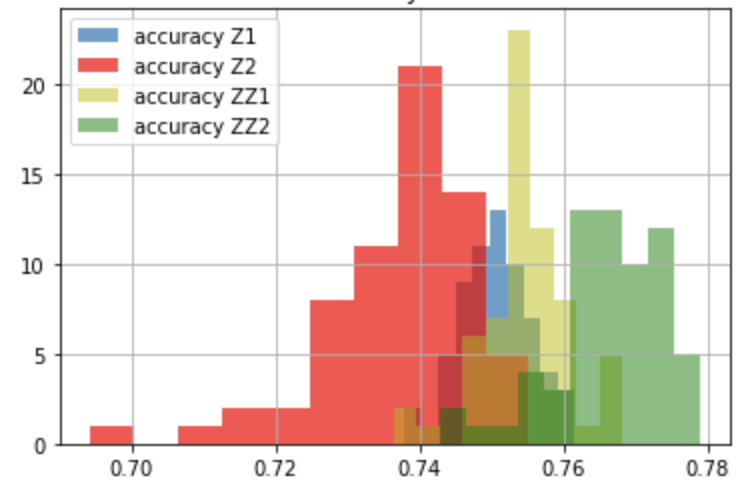}
  \includegraphics[width=0.24\textwidth]{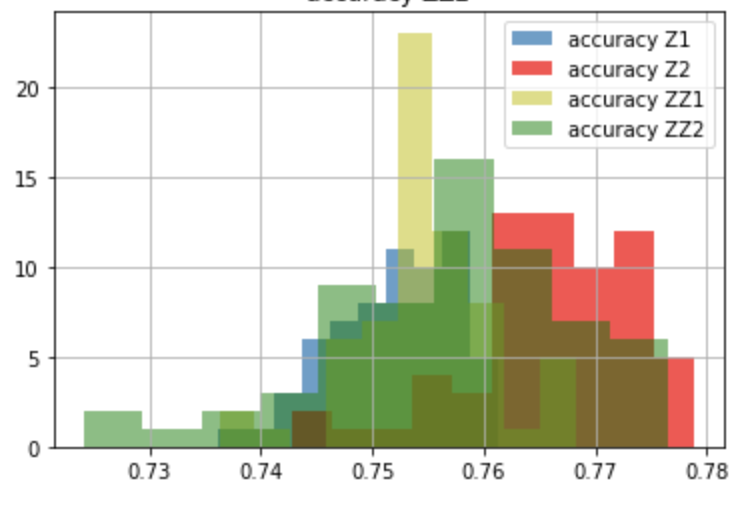}
  \includegraphics[width=0.24\textwidth]{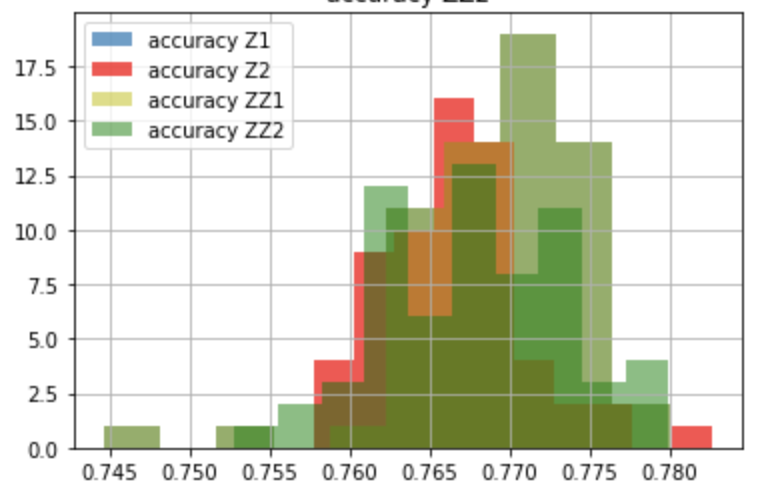}
  \caption{Figure shows the spread in the Accuracy Values when each feature combination is explored for best 3 features, best 4 features, best 5 features, best 6 features, respectively. This is done for each of the quantum feature map choices of Z map depth 1 and 2, as well as ZZ map depth 1 and 2.}
  \label{fig:spread}
\end{figure}

Figure \ref{fig:spread} shows the spread in accuracy values at each of these feature selection stages, where the best feature is selected at the maximal accuracy value for each of the Quantum Feature Map.

%\subsubsection{Best features based on State Vector Simulator}
%As briefly stated above, the procedure is performed initially under ideal condition, namely with a $state\_vector$ simulator. This is particular convenient because the first iterations are quite expensive in terms of calculation since the total number of permutation range from roughly half a million to thousands. This approach has been double checked in terms of performance and repeatability under noisy condition, targeting the execution of the algorithm on real hardware. 
%The search for features importance was performed using state vector simulator.

Once the best features were identified, we have run the final iteration and double-checked performance and repeatability under noisy condition, targeting the execution of the algorithm on real hardware.
Due to the abundance of replications and need of multiple trials we scripted the flow to allow for running the Quantum Instance which controls the transpilation and execution of a circuit via many different parameters, such as the backend, for simulation the noise model, basis gates, coupling map, etc., and is quite useful when wanting to run under different data input conditions: sample size, choice of data features. 
%under various set-up with the same data sample: 
{"training size": $1500$, "feature size": 7, "test size": $1000$, "order\_of\_expansion": "ZZ", "depth": $2$, "entanglement": "full", "alpha": $2.0$, "n\_shots": $8192$}.
\subsection{QSVM Results}

\begin{figure}[t]
  \centering
  \includegraphics[width=0.24\textwidth]{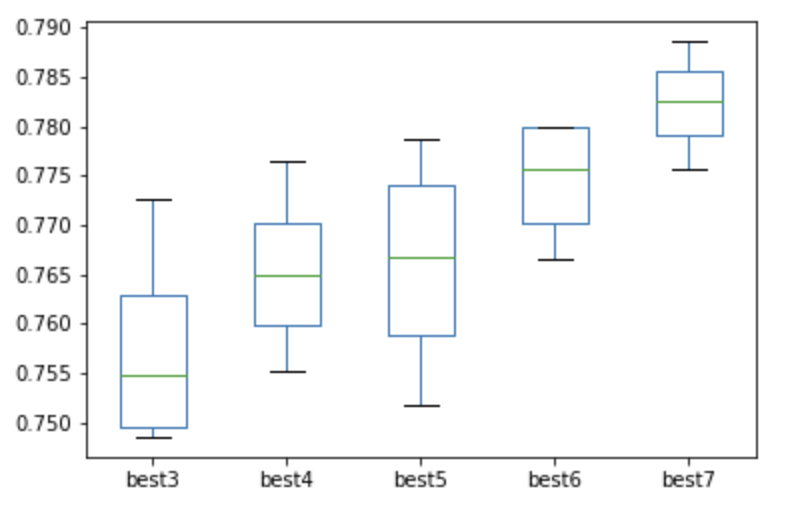}
  \includegraphics[width=0.24\textwidth]{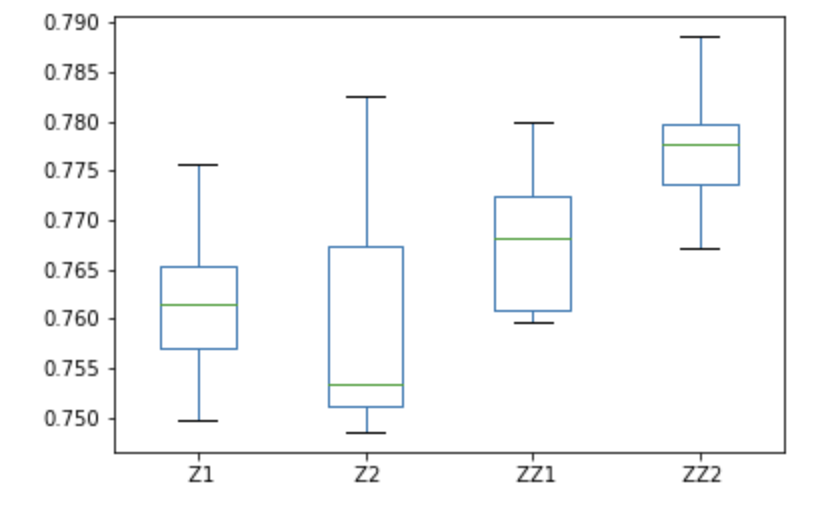}
  \caption{Figure a) shows the improvement of the accuracy as we add more features, while Figure b) shows the preference towards the entangled feature map for better accuracy.}
  \label{fig:FeatureEvolution}
\end{figure}

The search of best features is performed using Z and ZZ Quantum Feature Maps with depth $1$ and depth $2$. 
Figure \ref{fig:FeatureEvolution} provides an overview of the improvement observed in the accuracy when more features are added, as well as the preference towards the ZZ feature map with depth $2$. 
Feature maps with more entanglement performs better. The more entanglement a feature map uses the more difficult it is to simulate on classical hardware. As quantum hardware increases capacity in terms of qubits, the number if features that can be used will increase which may lead to even further improved performance. 

Another observation is that best features selected by this algorithm (Table \ref{tab:qsvm_7}) are different from the best features selected by the classical algorithms using same data sets, emphasizing a crucial role that feature exploration with QSVM plays to complement the feature space scanning. 
\begin{table}[ht]
\centering
 \small
\begin{tabular}{l|l|l|l}
\multicolumn{4}{c}{QSVM 6 Best Features} \\
ZZ depth 2 &  ZZ depth 1 & Z depth 1 & Z depth 2\\
\hline
\hline
%Cal\_Last4d\_TotAmt & Cal\_Last4d\_TotAmt & Cal\_Last4d\_TotAmt & Cal\_Last4d\_TotAmt % \\
%Response Code\_0 & SecuLevel\_0 & Response Code\_0 & Resp. Code\_51  \\
%Mer. Cntry Code\_HK  & Resp. Code\_0 & C\_4h\_Shop. &Resp. Code\_0\\
%\hline
%Accuracy= 0.772 & Accuracy=0.759 &Accuracy =0.749 & Accuracy=0.748 \\
%\hline 
%\hline
%+C\_4h\_ATM\>200  & +L\_CatMcc\_TravelAcc & +C\_POS\_Entry Mode\_other & %+C\_24h\_FREQ  \\
%\hline
%Accuracy= 0.776 & Accuracy=0.768 &Accuracy =0.761 & Accuracy=0.755 \\
%\hline 
%\hline
%+Merc.  Cat. Code\_5541 & +Amount& +C\_3h\_Declined & +Cardholder Cntry\_Italy \\
%Accuracy= 0.778 & Accuracy=0.772 & Accuracy=0.761 & Accuracy=0.751 \\
%\hline 
%+C\_4d\_US\_FREQ  & +Cal\_24h\_FREQ & +Amount & +E\_TimeLastATMwoInclCurr \\
%\hline
%Accuracy= 0.779 &Accuracy=0.779 & Accuracy=0.766 & Accuracy=0.771 \\
%\hline 
%\hline
%+Mer. Cntry Code\_GB & +C\_24h\_ATM\_Freq & +Amount & +Cal\_2dAgo\_FEQ\\
%\hline
%Accuracy= 0.788 & Accuracy=0.786 & Accuracy= 0.782 & Accuracy= 0.775 \\
F\_15 & F\_15 & F\_15 & F\_15  \\
F\_42 & F\_57 & F\_42 & F\_45  \\
F\_65  & F\_42& F\_10 &F\_42\\
\hline
Acc= 0.772 & Acc=0.759 &Acc =0.749 & Acc=0.748 \\
\hline 
\hline
+F\_9  & +F\_55 & +F\_31 & +F\_3  \\
\hline
Acc= 0.776 & Acc=0.768 &Acc =0.761 & Acc=0.755 \\
\hline 
\hline
+F\_36 & +F\_0& +F\_7   & +F\_48 \\
Acc= 0.778 & Acc=0.772 & Acc=0.761 & Acc=0.751 \\
\hline 
+F\_8 & +F\_3  & +F\_0   & +F\_19 \\
\hline
Acc= 0.779 &Acc=0.779 & Acc=0.766 & Acc=0.771 \\
\hline 
\hline
+F\_64   & +F\_2   & +F\_0 & +F\_13  \\
\hline
Acc= 0.788 & Acc=0.786 & Acc= 0.782 & Acc= 0.775 \\
\hline
\hline 
\end{tabular}
\caption{The 7 best features selected with QSVM feature selection algorithm under Various ZZ Map and depth selections. Each row cell corresponds to the first set of best selected features in increasing order of added features. The defining KPI is Accuracy. The best KPIs have been found for the ZZ depth 2.}
\label{tab:qsvm_7}
\end{table}

\begin{figure}[t]
  \centering
  \includegraphics[width=0.45\textwidth]{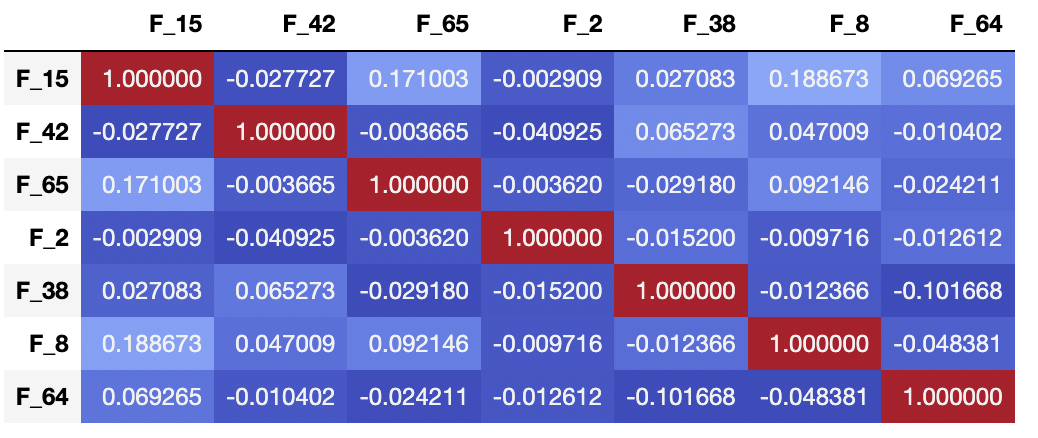}
  \caption{Correlation among best features selected by the QSVM method.}
  \label{fig:correl}
\end{figure}

Interesting observation is that the new algorithm has indeed identified features with least level of overlap, as shown in the correlation matrix in Table \ref{fig:correl}. This demonstrates that the choices are indeed viable. A reminder that most of the features have been engineered from same initial set of raw inputs, so identifying independent meaningful features is not a trivial find.

We observe that using this new feature selection by QSVM improves the outcome of the model when compared to use of QSVM with best classical features from XGBOOST and Random Forest from Table \ref{tab:imp}.  This comparison is performed using the exact same data sets with $1500$ records used for the training and $1000$ for testing.  Due to data under-sampling, we have used $5$ random  trials to minimize bias. The average KPIs for accuracy and ACU are reported in the Table \ref{tab:test_features}.  

\begin{table}[ht]
    \centering
    \small
    \begin{tabular}{l|l|l|l}
    KPI QSVM & w/ XGB bf & w/ RF bf. & w/ QSVM bf\\
    \hline
    \hline 
Accuracy (Test)& $0.76 \pm 0.01$ & $0.76 \pm 0.01$  & $0.78 \pm 0.01$ \\
AUC(Test)& $0.81 \pm 0.01$ &   $0.81 \pm 0.01$ & $ 0.81 \pm 0.01$ \\
\hline
\hline
    \end{tabular}
    \caption{Accuracy and AUC results for QSVM using XGBoost and Random Forest best feature selection vs QSVM best feature selections, based on Balanced Data Set and using $6$ trials.}
    \label{tab:test_features}
\end{table}

\begin{figure}[t]
  \centering
  \includegraphics[width=0.4\textwidth]{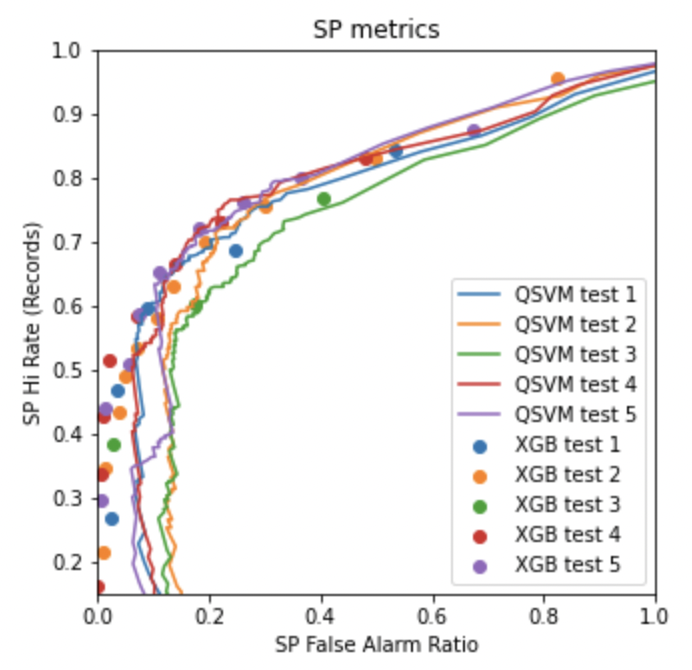}
  \includegraphics[width=0.4\textwidth]{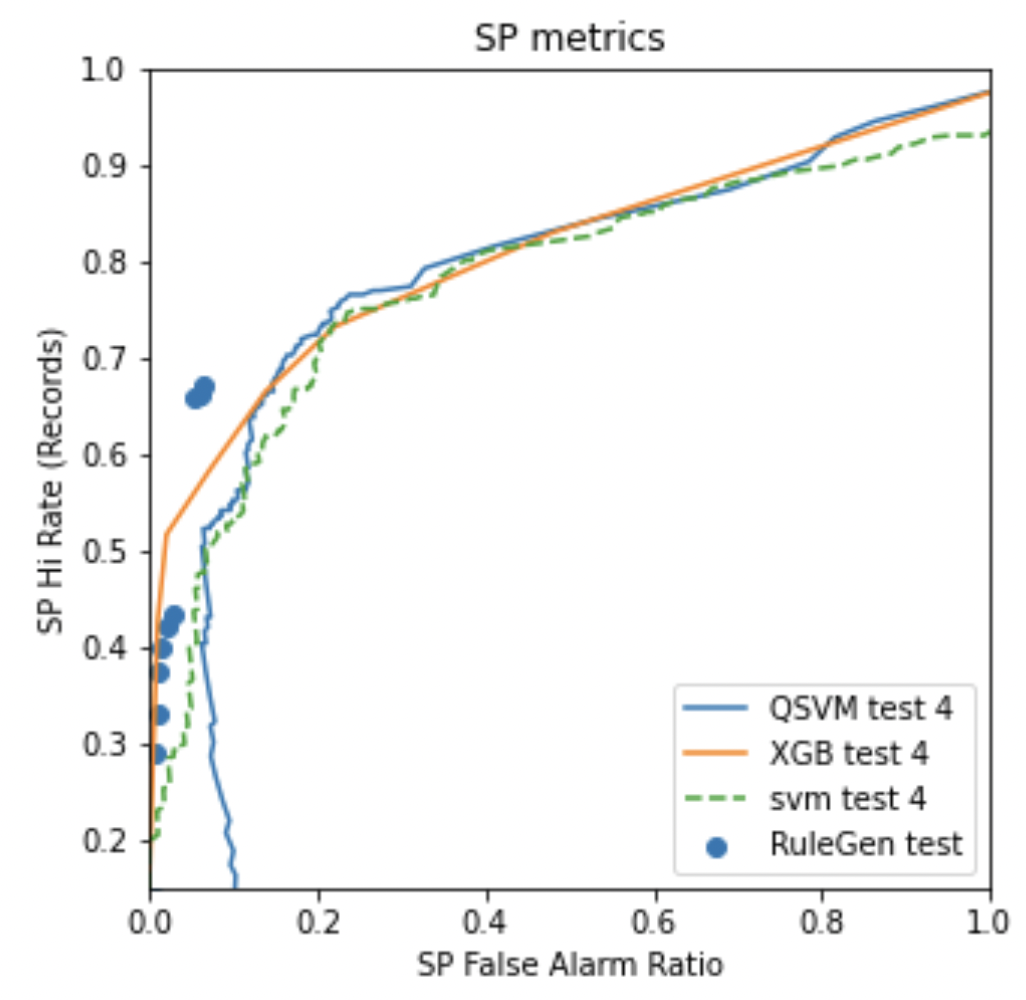}
  \caption{Figure shows the modified ROC* Curve for QSVM using state vector simulator.}
  \label{fig:FeatureEvolutionROC}
\end{figure}

\begin{figure}[t]
  \centering
  \includegraphics[width=0.4\textwidth]{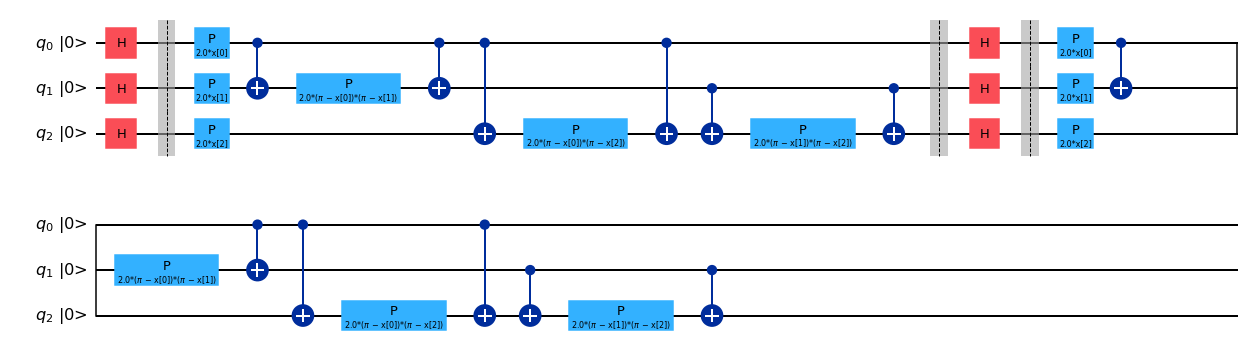}
  \caption{Example of a QSVM transpiled job circuit for 3 features.}
  \label{fig:depth}
\end{figure}

We ran the model using different backends available on the IBM Quantum platform.  As explained, the workflow was to start with the ideal simulator (state\_vector).  We used this simulator for running the algorithm to determine best features. Once best features were found, we repeated the run on qasm-simulator. The standard deviation is estimated from repeating the run on 6 trials. The KPI that we recorded for the case without accounting for noise is close to the one we encountered with the state vector simulator. 
While many opportunities exist to use quantum computing systems, there are many facets that go along with it such as software, cloud access, benchmarking, and error correction and mitigation \cite{corcoles2019challenges,wood2020special}. Since the current state-of-the-art for these systems are still considered noisy, we wanted to run our simulation tests with noise models which are based off real quantum systems \cite{georgopoulos2021modeling}.  We have used the noise source analyzed for the ibm\_cairo backend and re-run the QSVM and enabled the readout measurement error mitigation flag.  
When using noisy simulation, the optimal circuit transpilation pipeline in Qiskit is provided with the parameter optimization\_level set to $3$ which selects a candidate initial\_layout and SWAP mapping using the Sabre layout and routing method \cite{DBLP:journals/corr/abs-1809-02573}, and performs the most 1Q and 2Q gate optimizations. 

The circuit depth produced by the $7$ best features is around $70$, an illustration of a circuit depth diagram produced by $3$ features is provided in Fig. \ref{fig:depth}. Reducing the depth of the circuit and optimize it for use on the hardware is a study for future work that is in plans.
% selects a good candidate initial\_layout and SWAP mapping using the Sabre layout and routing method (arXiv:1809.02573), and performs the most 1Q and 2Q gate optimizations. Because this method is stochastic, finding a good candidate SWAP mapping can involve mapping the circuit several times and selecting the one with the fewest entangling gates (lowest depth).

%We have also attempted to run on the quantum hardware. We ran on different IBM backends:  ibm\_kolkata and ibm\_mumbai, $27$ qubits systems and 128 Quantum Volume, with average times for T1 and T2 of order of $100 \mu s $ and average CNOT Errors of $8.2\times 10^{-2}$. The initial layout has been hand picked for least noisy qubits according to single and 2-qubit gate errors. We ran the circuit with optimization level $3$ for the transpiler, and we tested the impact of enabling the measurement error mitigation (m.e.m.) flag.  However, as results show  in Table \ref{tab:overview}, the model returns no discrimination power in performing KPIs. 
 
% To do:
% \begin{itemize}
%     \item plans to improve depth ($\gamma$  factor) \red{Noelle}
%     \item circuit knitting, embedded 
% \end{itemize}

\begin{table}[ht]
    \centering
    \small
    \begin{tabular}{l|l|l}
    backend & Accuracy & AUC \\ 
    \hline
    \hline
    statevector sim. & $0.78\pm 0.01$ & $0.81 \pm 0.01$\\
    qasm sim. w/o noise & $0.77 \pm 0.03$ & $0.79 \pm 0.05$ \\
    qasm sim. w/ noise & $0.55\pm 0.10$& $0.74\pm 0.14$ \\ 
    \hline
%   ibm\_mumbai  & $0.52$ & $0.44$ \\
%   ibm\_kolkata  &  $0.52$ &  $0.51$\\
%   ibm\_kolkata with m.e.m.  &  $0.52$&  $0.54$\\
    \hline
    \hline
    \end{tabular}
    \caption{KPIs for test samples when running QSVM on different backends: state vector simulator, qasm simulator with and without noise, IBM quantum systems with and wirhout m.e.m. enabled.}
    \label{tab:overview}
\end{table}

\section{Mixed Quantum-Classical Method For Fraud Detection}%Noelle

In this paper we are exploring the complementarity to classical ML provided by QSVM.

Although quantum computing has been proven to speed up some types of problems  \cite{provenspeedup}, the existent technology allows only a limited number of qubits and gate operations. Therefore, we employ a hybrid classical / quantum solution where
data and function learning are classical, while the classification algorithm is quantum. In this paper we use Quantum Support Vector Machine, where the quantum
computer is only used once to compute the kernel matrix for
the training set. The optimal separating hyperplane and the
support vectors can be obtained efficiently in the training set
size by solving the conventional dual optimization problem on
a classical computer.

\subsection{Approach}
In order to exploit the complementarity of the quantum and classical algorithms to increase classification performance, we wish to discern those transactions or data points for which the two algorithms disagree in the classification. When the two algorithms disagree, we wish to predict which is correct. In order to accomplish this, we trained both the quantum and classical algorithms on the train sub-portion of the balanced dataset. When the two algorithms disagreed on the label of a given transaction in the training set, the transaction was noted. These transactions, a subset of the training data of the balanced dataset, formed an additional dataset on which a metaclassifier was subsequently trained. The metaclassifier may take as features any of the features from the dataset and may take on a number of forms. In practice, because of the size of dataset, the number of training datapoints on which the classifiers disagreed was limited so a simple metaclassifier performed best. In the case of the fully optimized XGBoost a classical SVM was used as the metaclassifier. In the case of the XGBoost without optimization, a logistic regression was used.
\subsection{Results and Comparisons}

\begin{table}[ht]
    \centering
    \small
    \begin{tabular}{l|l|l|l|l|l}
  Model & Acc. &  CI  & Std &N\_\tiny{trials} & N\tiny {best perf. features}\\ \hline
 XGBoost & 77.7\% &  0.3\% &  0.5\% &  10 &  9\\
 XGBoost - opt.&  80.0\% & 0.5\% & 0.9 \% & 10 & 37\\
 QSVM - ZZ & 78.8\% & 0.4\% & 0.7\% & 10& 7\\

   \end{tabular}
    \caption{Comparison of Model Accuracy on Balanced Dataset - Separate Models}
    \label{tab:xgbimpsep}
\end{table}

    \begin{table}[ht]
    \centering
    \small
    \begin{tabular}{l|l|l|l|l}
  Model & Acc. &  CI  & Std &N\_trials\\\hline
 QSVM + XGBoost & 79.9\% & 0.4\% & 0.7\% & 10 \\ 
 QSVM + XGBoost- opt.&  81.0 &0.3\%  & 0.5\% & 10 \\

   \end{tabular}
    \caption{Comparison of Model Accuracy on Balanced Dataset - Mixed Models}
    \label{tab:xgbimpmixed}
\end{table}

 \begin{figure}[t]
  \centering
  \includegraphics[width=0.45\textwidth]{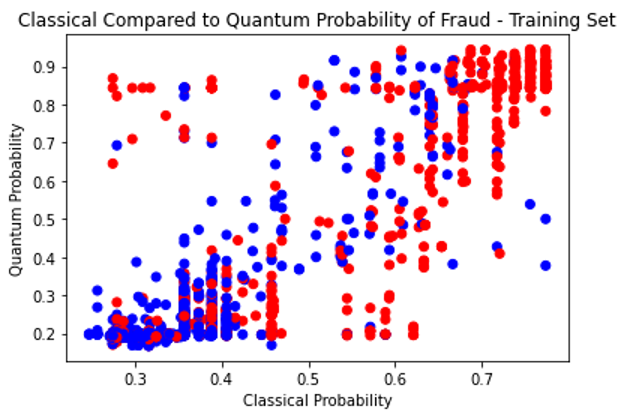}
  \includegraphics[width=0.45\textwidth]{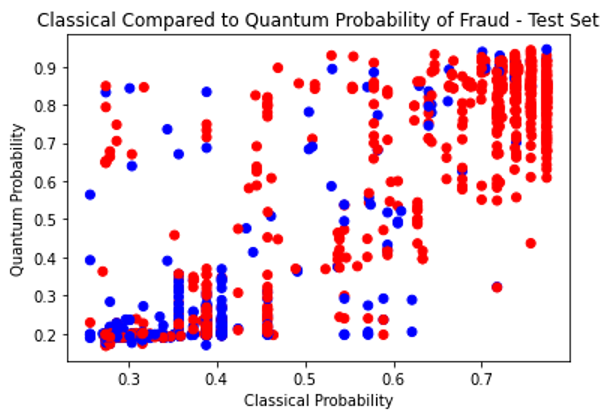}
  \caption{Figure shows complementarity of the Classical and Quantum decisions. Red  Dots are fraudulent transactions, Blue dots are genuine transactions. Although similar Performance, different label Assignments. The position on the x axis represents the probability of fraud predicted by the classical algorithm; dots with high probability are predicted to be fraud by the classical algorithm. The position on the y axis represents the probability of fraud predicted by the quantum algorithm; dots with high probability are predicted to be fraud by the quantum algorithm.}
  \label{fig:complementarity}
\end{figure}
While the performance of the classical and quantum algorithms was similar, the actual predictions can vary for specific datapoints, yielding complementary results (see Figure \ref{fig:complementarity}).  Classifications disagree on 5.2\% of Training Data,  and  5.5\% of Test Data, with threshold of 0.5.
On the diagonal the quantum and classical models agree. On the off-diagonal they disagree. This shows that different relationships are detected  by the quantum and classical  models- complementarity. We exploit this complementarity to increase performance using a metaclassifier which determines which algorithm to "believe", given the surrounding circumstances as expressed by the features of a given transaction. 
\begin{figure}[t]
  \centering
  \includegraphics[width=0.48\textwidth]{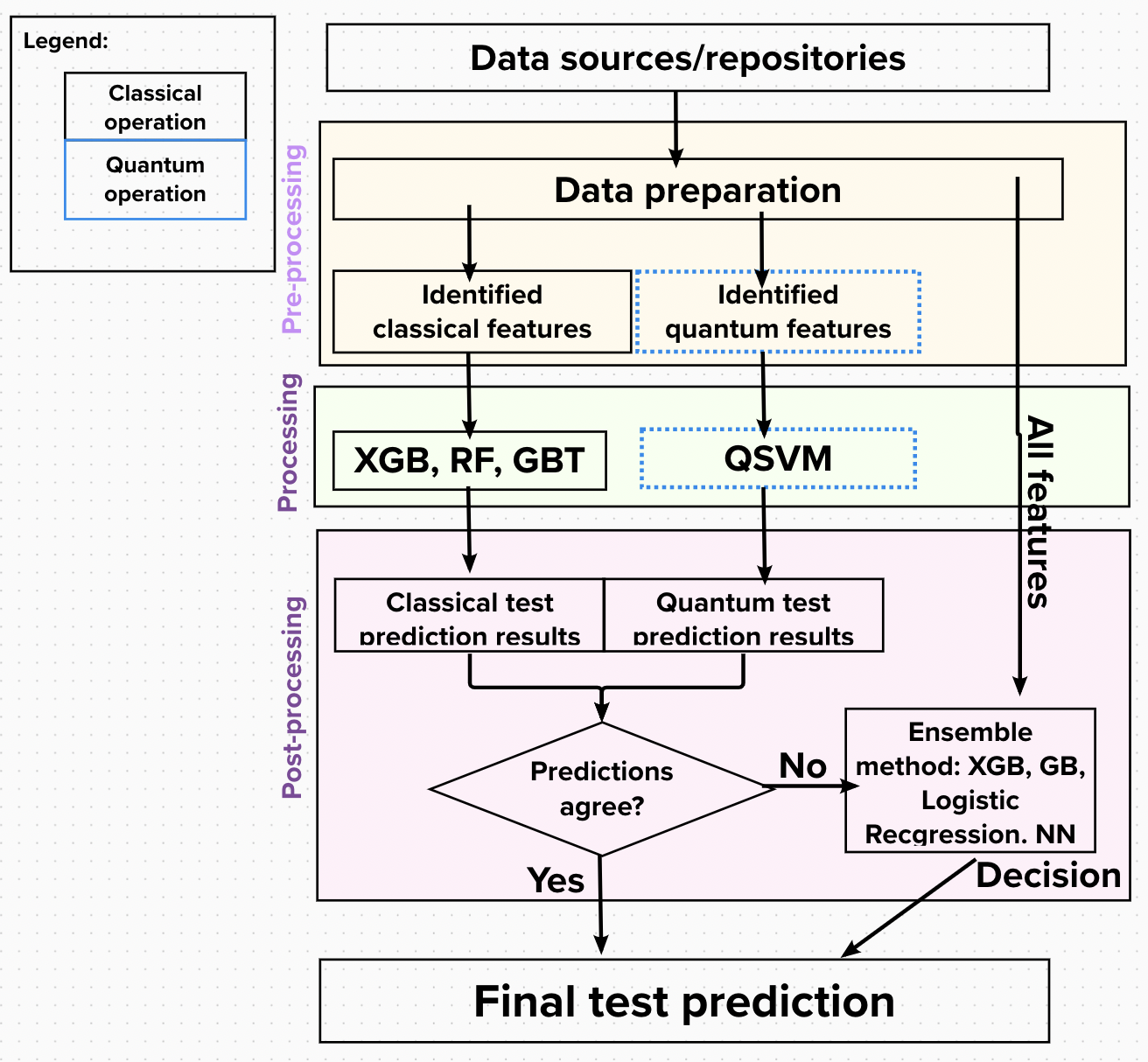}
  \caption{The Flow chart of how to combine the Classical and Quantum Approached into a single decision. When the quantum and classical algorithms disagree – a Metaclassifer predicts which one is correct.}
  \label{fig:mixflow}
\end{figure}

%Quantum algorithm can be prepared with python, with circuits defined with python too.
Using IBM Qiskit simulator, we have successfully employed Quantum Support Vector Machine method and we have measured the accuracy and AUC for different use cases.
%\section{Future Work - Noelle}
%While using quantum computing to create better performing models in itself can instantly save billions of dollars of fraud losses per year for consumers, the true value of quantum computing would come from applying quantum computing to feature discovery. In payment fraud prevention, the true power of a model comes from the quality of the features it is provided with. And these features need to identify complex matters. For example, a certain sequence of financial and non-financial transactions can indicate identity theft. Or that a card is “tested” for not being blocked before an expensive fraudulent purchase is made. Or transaction signals of account takeover. Current feature discovery methods cannot 
%deal well with the complexity and combinatorial aspect of this task. Industry’s best practice still mostly relies on the analytical mind of experienced domain experts. The authors believe that the ability of quantum computing to deal well with the aspects of complexity and combinatorial could open the possibility of automated generation of high performing behavioral features that will result in better performing models and hence substantial fraud loss savings while at the same time making payments more frictionless for customers.

%You can use the standard commands for equations.
%\begin{align}
%  \label{emc}
%  E &= m\,c^2\\
%  a^2 + b^2 &= c^2\\
%  H\,|\psi\rangle &= E\,|\psi\rangle\\
%  (\openone \otimes A)\,(B \otimes \openone) &= A %\otimes B
%\end{align}

\section{Summary}

Classical Machine Learning algorithms are currently state of art for predicting fraud in transactions.  Quantum machine learning can provide a complementary support on this, exploiting enhanced feature space to encode historical data. In this work we propose a novel approach to maximize a quantum classifier performance in terms of accuracy of prediction, but other KPI can be explored as well.
The method is called Quantum Feature Importance Selection algorithm:  
 using Quantum enhanced Support Vector Machine we are able to select most relevant features for the quantum classifier for an increasing number of selected features. In this case we also noted that Quantum Feature Map that makes use of more entanglement provides systematically better KPIs.
The whole workflow requires quite an intensive care for data pre-processing from data type considerations to under-sampling techniques before moving to the quantum part.  We found that quantum classifiers can identify different types of patterns in the data that are difficult for classical machine learning algorithms to detect while being complimentary to classical machine learning algorithms.
We also define a mixed quantum-classical ensemble method that can help businesses strike a better balance between false positives and false negatives and improve the KPI of the final model. The result presented are obtained on a simulated quantum computer and the extension of this work to real hardware implementation will be collected in a different manuscript.

\section*{Acknowledgment}

We acknowledge the use of IBM Quantum cloud platform for this work. The views expressed are those of the authors, and do not reflect the official policy or position of IBM or the IBM Quantum team.
M.G. is supported by CERN Quantum Technology Initiative.

\bibliographystyle{IEEEtran}
\bibliography{references.bib}

\EOD

\end{document}